\documentclass{aa} 
\usepackage{graphicx} 
\usepackage{txfonts} 
\newcommand{\mum}{\,$\mu$m} 
\begin{document}

\title{Observational evidence for the presence of PAHs in distant \\
Luminous Infrared Galaxies using ISO and {\it Spitzer}}
 
\titlerunning{Evidence for PAHs in LIRGs w/ ISO and {\it Spitzer}}
 
\author{\bf D. Elbaz\inst{1}, 
E. ~Le Floc'h\inst{2}, 
H. ~Dole\inst{3}, 
D. ~Marcillac\inst{1}\thanks{Based on observations collected with
the Spitzer Space Telescope (which is operated by the Jet Propulsion
Laboratory, California Institute of Technology under NASA contract
1407) and on observations with ISO (an ESA project with instruments
funded by ESA Member States (especially the PI countries: France,
Germany, the Netherlands and the United Kingdom) with the
participation of ISAS and NASA)}
} 
\offprints{D. Elbaz, email: delbaz@cea.fr} 
 
\institute{DSM/DAPNIA/Service d'Astrophysique, CEA/SACLAY, 91191 Gif-sur-Yvette Cedex, France 
\and 
Steward Observatory, University of Arizona, 933 N Cherry Ave, 85721
Tucson, AZ, USA
\and 
Institut d'Astrophysique Spatiale, b\^at 121, Universit\'e Paris-Sud,
F-91405 Orsay Cedex, France
} 

\date{Received: ; Accepted: } 
 
%\titlerunning{Detection of PAHs with SPITZER} 
%\authorrunning{Elbaz et al.} 
 
%---------------------------------------------------------------------- 
\abstract{We present ISOCAM~15\mum~and MIPS~24\mum~photometry of a sample
of 16 distant Luminous Infrared
Galaxies (LIRGs) characterized by a
median luminosity L$_{\rm IR} \sim$ 2$\times$10$^{11}$ L$_{\odot}$ and
redshift $z = 0.7$ (distributed from $z = 0.1$ to 1.2). While some sources
display 24/15\mum~flux ratios also
 consistent with a featureless continuum dominating their mid-infrared (MIR)
spectral energy distributions (SEDs), 
the presence of prominent emission features such as the Polycyclic Aromatic Hydrocarbons 
is clearly required to explain the observed colors for more than half of the sample.
As a result,  a general good agreement is observed between the data and predictions
from the local starburst-dominated SEDs that have been used so far to constrain 
IR galaxy evolution. This is consistent with 
the star-forming nature of LIRGs derived from
previous works, even though our approach cannot rule out
the dominance of an AGN in some cases.
Our study  also supports  the possibility of tracing the 
total IR luminosity of distant galaxies
(up to $z\sim$ 1) from their MIR emission.
} 
\maketitle 
%---------------------------------------------------------------------- 
\section{Introduction} 
%---------------------------------------------------------------------- 
The excess of faint sources in the deep extragalactic surveys
performed in the MIR to sub-millimeter with their associated redshift
distributions when available, together with the shape and intensity of
the cosmic IR background (CIRB, Puget et al. 1996), which measures the
extragalactic IR light radiated above 3\mum~and is at least equal to
the UV to near-IR extragalactic background light (Gispert et al. 2000;
Hauser \& Dwek 2001), suggest that a large fraction of the UV
radiation of young stars was reprocessed by dust in the mid to far IR
range over the Hubble time. The cosmic star formation history- the
star formation rate (SFR) per unit comoving volume as a function of
redshift- would be dominated by intense star formation phases, during
which the bulk of the UV light is reprocessed by dust in the IR with a
rapid decline of the SFR density since $z\sim$ 1 and a flat or
possibly slow increase from $z\sim$ 5 to 1 (Chary \& Elbaz 2001,
Lagache et al. 2004 and references
therein). These models called ``backward evolution models'' because
they evolve the local luminosity function in the MIR as a function of
redshift, assume that the shapes of distant SEDs in the MIR remain
similar to the ones observed in local galaxies and are directly
related to the total IR luminosity.

The goal of the present study is to check the validity of this
assumption up to $z\sim$ 1 by combining data from ISOCAM (Cesarsky
et al. 1996a), the MIR camera onboard the Infrared Space Observatory
(ISO, Kessler et al. 1996), and MIPS, the Multiband Imaging Photometer
for {\it Spitzer} (Rieke et al. 2004) onboard the {\it Spitzer} Space
Telescope (Werner et al. 2004). Studying the MIR shape of distant
galaxies is particularly relevant since direct FIR observations are
rapidly confusion limited and strongly affected by moderate
sensitivity limits. On the other side of the peak of the IR SED,
sub-millimeter observations are presently limited to the detection of
ultra-luminous IR galaxies (ULIRGs) above $z\sim$ 2 (Chapman et al.
2003). In the near future, before the launch of Herschel and the
advent of ALMA, the best constraints on the evolution of luminous IR
galaxies (LIRGs), i.e. galaxies with L$_{\rm IR} \geq $ 10$^{11}$
L$_{\odot}$, as a function of redshift and on their role in the global
evolution of galaxies will exclusively come from the MIR. Around $z=$
0.7, LIRGs are too faint to be measured through direct spectroscopy in
the MIR with the infrared spectrograph (IRS, Houck et al. 2004a) onboard {\it
Spitzer}. The only remaining technique to contrain the MIR shape of
distant LIRGs is therefore to combine MIR measurements for LIRGs
located at different redshifts, hence spanning a larger range of
rest-frame wavelengths. In the present study, a redshift range of $z=$
0.1-1.2 allowed us to study the 5 to 25\mum~rest-frame part of the MIR
SEDs. In this wavelength range (see Genzel \& Cesarsky 2000), the SED
is dominated by the combination of broad emission lines, generally
interpreted as due to polycyclic aromatic hydrocarbons (PAHs), and of
the continuum emission of stochastically heated ``very small grains''
of dust transiently heated to temperatures of the order of 200 K. The
thermal emission of ``big grains'' of dust heated to $\sim$ 40 K also
contributes partly to the MIR emission but peaks in the FIR (between
80 and 100\mum~ typically) and contains the bulk of the luminosity
radiated by galaxies above 3\mum. Finally, hot dust emission due to
dust heated by an active galaxy nucleus (AGN) can also contribute to
and sometimes even dominate the MIR spectrum of a galaxy. Before
deriving the IR luminosity for a galaxy that will be used to compute
its SFR, one must start by making sure that its MIR SED is not
polluted by the AGN emission. The presence of PAHs as well as a rapid
decline of the continuum emission below 5\mum~strongly suggests a star
formation origin for the emission (Genzel \& Cesarsky 2000, and
references therein). Local LIRGs are rarely affected by an AGN,
contrary to ULIRGs above 10$^{12.3}$ L$_{\odot}$. Using a combination
of template SEDs and deep X-ray surveys with XMM-Newton and Chandra,
Fadda et al. (2002) derived an upper limit to the contribution of an
AGN to the MIR light radiated by ISOCAM selected LIRGs up to $z\sim$ 1
of 20\,\%.

The MIR emission of local star forming galaxies- at 6.75, 12 and
15\mum- was proven to correlate, with some scatter, with the total IR
one, i.e. from 8 to 1000\mum, which is largely dominated by the FIR
component (Chary \& Elbaz 2001, Elbaz et al. 2002). However, the
validity of these correlations in the more distant universe has not
yet been established. Galaxies were less metal rich in the past. Their
shape and compacity evolving with time might also affect their SEDs.
Local galaxies SEDs do present some variations in the MIR as a
function of metallicity. The metal-deficient (Z=Z$_{\odot}$/41) blue
compact dwarf galaxy SBS 0335-052, for example, is bright in the MIR
range but still does not show any sign of the presence of the PAH
features which the authors interpret as an effect of the destruction
of their carriers by the very high UV energy density (Thuan, Sauvage
\& Madden 1999, Houck et al. 2004b).

Throughout this paper, we will assume H$_o$= 70 km s$^{-1}$ 
 Mpc$^{-1}$, $\Omega_{\rm matter}$= 0.3 and $\Omega_{\Lambda}= 0.7$. 
%
%---------------------------------------------------------------------- 
\begin{figure}[t] 
\resizebox{8.5cm}{!}{\includegraphics{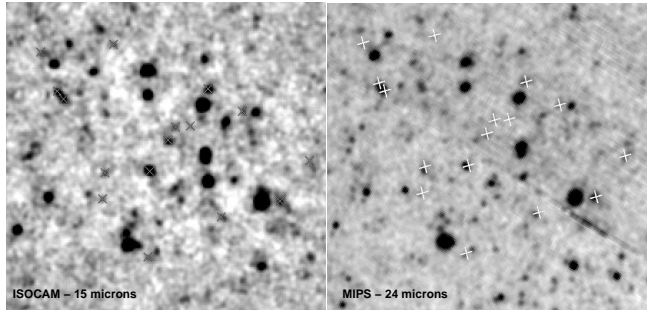}} 
\caption{{\it Left:} ISOCAM 15\mum~image of the Marano FIRBACK field
(central part of the ultra-deep survey, 7$\arcmin\times$7$\arcmin$)
down to an 80\,\% completeness limit of 140\,$\mu$Jy.  Black X
represent the positions of the common sources with MIPS plotted in the
Fig.~\ref{FIG:all_seds}.  {\it Right:} MIPS 24\mum~image of the same
field down to an 80\,\% completeness limit of 170\,$\mu$Jy. White X
represent the position of ISOCAM sources.
}
\label{FIG:cam_mips} 
\end{figure} 
%---------------------------------------------------------------------- 
%---------------------------------------------------------------------- 
\begin{figure} 
%\resizebox{8.5cm}{!}{\includegraphics{multi_z_ordered_SEDlum_compact.ps}} 
\resizebox{8cm}{!}{\includegraphics{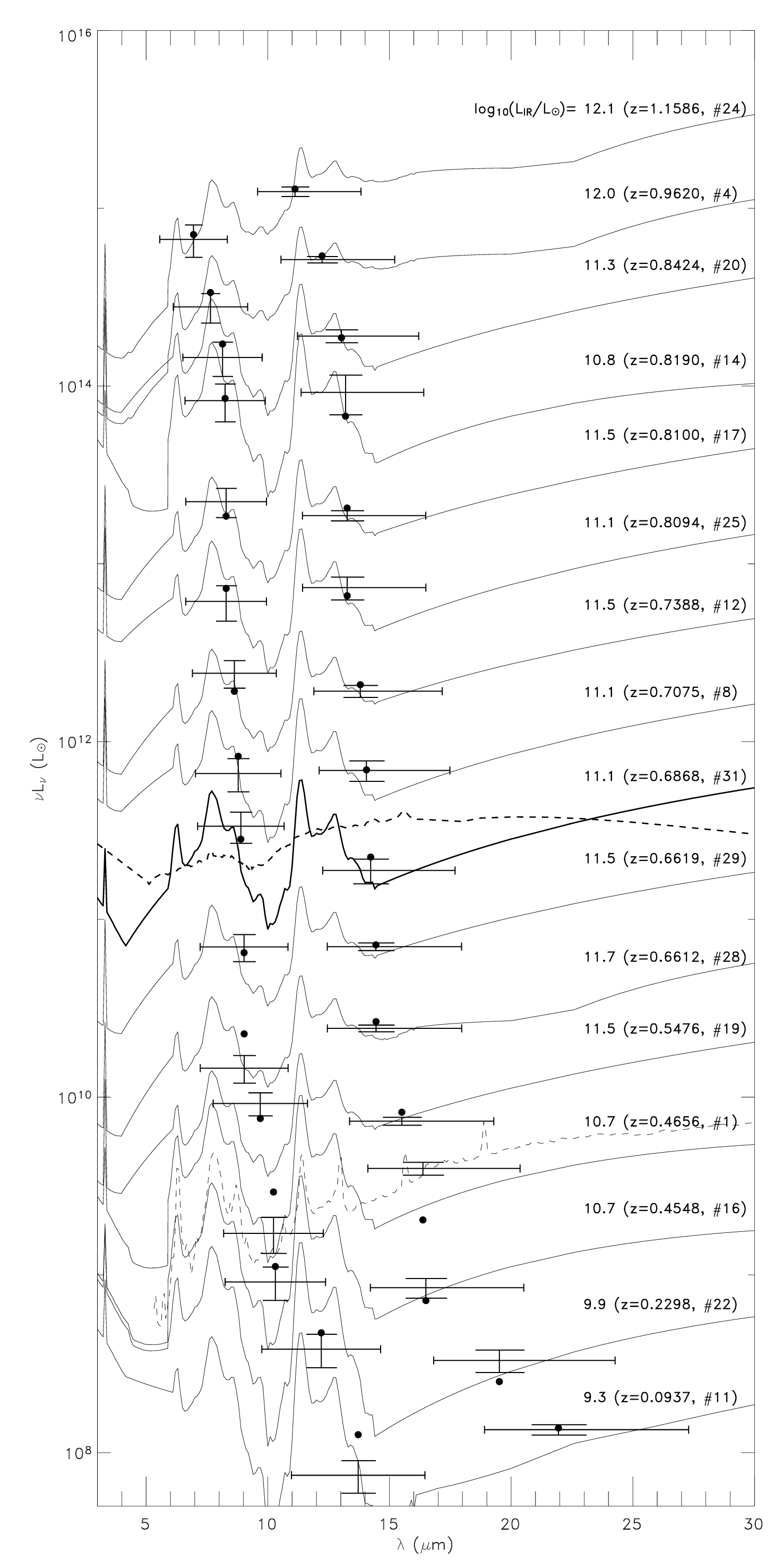}} 
\caption{All SEDs sorted (from top to bottom) by decreasing redshift.
The SEDs are shifted by an arbitrary offset in $\nu$L$_{\nu}$ for
visibility; wavelengths are rest-frame. ISOCAM 15\mum~and MIPS 24\mum~
luminosities are reported with the filter bandwidth and the 1-$\sigma$
uncertainty. The label indicates the logarithm of the IR luminosity as
well as the redshift and ID of each source, e.g. ID \#4 is the galaxy
UDSF04 in Liang et al. (2004). The filled circles are the luminosities
that would be measured in both filters for the plotted SED from the
library of template SEDs of Chary \& Elbaz (2001). Bold dashed line on
gal.\#31: SED of the AGN NGC 1068 normalized to best fit the measured
15 and 24\mum~luminosities of the galaxy. Dashed line on gal.\#1: SED
of NGC 7714 (Brandl et al. 2004).}
\label{FIG:all_seds} 
\end{figure} 
%---------------------------------------------------------------------- 
%---------------------------------------------------------------------- 
\begin{figure} 
%\resizebox{\hsize}{!}{\includegraphics{elbaz_F3.ps}} 
\resizebox{8.15cm}{!}{\includegraphics{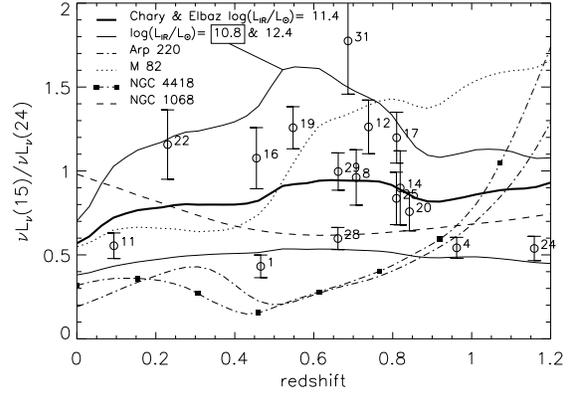}} 
\caption{The observed colors $\nu L_\nu(15)/\nu L_\nu(24)$ derived
from our sample (open circles and vertical error bars, with numbers
refering to IDs from Fig.~\ref{FIG:all_seds}), compared with the
predictions for different object types such as the Seyfert~2 prototype
NGC~1068, two enshrouded IR-luminous systems with deep silicate
absorption (Arp\,220 and NGC\,4418), M82 (log(L$_{\rm
IR}$/L$_{\odot}$)=10.6) and three template SEDs from Chary \& Elbaz
(2001, plain lines; the lowest line corresponds to log(L$_{\rm
IR}$/L$_{\odot}$)=12.4).  }\label{FIG:mean_sed}
\end{figure} 
%---------------------------------------------------------------------- 

%---------------------------------------------------------------------- 
%---------------------------------------------------------------------- 
%---------------------------------------------------------------------- 
\section{Sample selection and data reduction}
\label{SECTION:obs}
%---------------------------------------------------------------------- 
%---------------------------------------------------------------------- 
A 30'$\times$30' field was imaged by ISOCAM at 15\mum~in the Marano
FIRBACK region centered on (3h13m10s,-55$^o$03'39''), about two thirds
of which is covered by a larger MIPS-24\mum~image of
1.4$^o\times0.7^o$ with a 80\,\% completeness limit of
170\,$\mu$Jy. The central part of this field (7'$\times$7', UDSF for
``ultra-deep survey firback'') was covered down to an 80\,\%
completeness limit of 140\,$\mu$Jy at 15\mum~with ISOCAM, a depth
similar to the MIPS one (see Fig.~\ref{FIG:cam_mips}). The
ISOCAM-15\mum~UDSF belongs to the ISOCAM Guaranteed Time Extragalactic
Surveys (IGTES, Elbaz et al. 1999). Six overlapping mosaics were
performed in the micro-scanning mode of ISOCAM for a total of 2.7
hours of integration per sky pixel and leading to a final projected
map with 2$\arcsec$ pixels. The data were reduced using PRETI (Starck
et al. 1996) and the transient correction as well as the assessment of
the completeness limit were computed by Monte Carlo simulations with
fake sources of known flux densities in the real images.

The 24\mum~observations were performed during the MIPS commissioning
phase (IOC/SV) in November 2003. The scan map AOT was used, with an
half-array overlap to cover about 1300 Sq. Arcmin with high redundancy
(20) and to get an integration time per sky pixel of about 230s
(Papovich et al. 2004). The data were reduced using the {\it Spitzer}
Science Center Pipeline and the BCD products (Basic Calibrated Data, Pipeline
version S10.0.3)
were downloaded from the {\it Spitzer}
archive\footnote{http://archive.spitzer.caltech.edu, PID: 718}. 
 PSF-fitting photometry was performed using DAOPHOT
(Stetson 1987) with IRAF\footnote{IRAF is distributed by the National
Optical Astronomy Observatories, which are operated by the Association
of Universities for Research in Astronomy, Inc., under the cooperative
agreement with the National Science Foundation}.

For the present study, we used a sample of 16 galaxies detected with
both ISOCAM and MIPS at 15 and 24\mum~respectively and identified with
crosses in Fig.~\ref{FIG:cam_mips}. This set of galaxies was extracted
from a larger sample of ISOCAM selected galaxies for which medium
resolution (R=1200) VLT-FORS2 spectra were obtained (Liang et al. 2004)
in the ``Marano FIRBACK'' field. Oxygen abundances were obtained for
half of them with values of 12+log(O/H) ranging from 8.4 to 9. With a
median stellar mass of 5$\times$10$^{10}$ M$_{\odot}$ and redshift of $z\sim$ 0.7, they are among the most massive
galaxies in this redshift range but exhibit
metallicities about twice smaller than their nearby counterparts
suggesting that they will produce about half of their metals from
$z=$1 to 0 (Liang et al. 2004). Finally, it must be
noted that even though no direct FIR measurement exists for this
sample of galaxies at present, there is indirect evidence that the
MIR-FIR correlations remain valid up to $z\sim$ 1.  Indeed, MIR and
radio measurements provide consistent predictions for the FIR
luminosity of distant galaxies up to $z\sim$ 1 (Elbaz et al. 2002,
 Appleton et al. 2004).

%---------------------------------------------------------------------- 
%---------------------------------------------------------------------- 
%---------------------------------------------------------------------- 
%\section{Spectral Energy Distributions} 
\section{Discussion} 
\label{SECTION:sed}
%---------------------------------------------------------------------- 
%---------------------------------------------------------------------- 
%---------------------------------------------------------------------- 
Fig.~\ref{FIG:all_seds} presents the 15 and 24\mum~luminosities of the
16 galaxies detected with ISOCAM and MIPS in the UDSF fitted by a
library of template SEDs from Chary \& Elbaz (2001). This library of
100 template SEDs, from log(L$_{\rm IR}$/L$_{\odot}$)= 8.5 to 13.5,
was generated between 0.1 and 1000\mum~to reproduce the trend observed
for local galaxies between MIR and FIR luminosities. The MIR
(4$-$20\mum) part was produced by interpolating between the ISOCAM CVF
spectra of four prototypical galaxies (Arp220, NGC6090, M82 and M51),
hence all SEDs contain PAH features.  Because the 6.75, 12 and
15\mum~luminosities of a local galaxy correlate with its total IR
luminosity, $L_{\rm IR}$, it is possible to use a single MIR
measurement to derive $L_{\rm IR}$, and therefore a SFR, for nearby
galaxies. In ``backward evolution models'', such as Chary \& Elbaz
(2001), the 15\mum~luminosity function is evolved as a function of
redshift to reproduce the observe galaxy counts and the CIRB, assuming
that a template SED is uniquely determined by a given
15\mum~luminosity. To fill the gaps between the observed 6.75, 12 and
15\mum~correlations, the templates interpolated for all total IR
luminosities were used. We used the same technique to derive $L_{\rm
IR}^{15}$ from $L_{\rm 15}$ (measured with ISOCAM), $L_{\rm IR}^{24}$
from $L_{\rm 24}$ (measured with MIPS) and $L_{\rm IR}^{15-24}$ from
the combination of $L_{\rm 15}$ and $L_{\rm 24}$.  First, the 100
template SEDs were redshifted at the redshift of a given galaxy among
the 16 and a flux density was determined at 15\mum~using the filter
response of ISOCAM. Second, the SED which 15\mum~flux density was the
closest to the observed one was selected and normalized to reach the
observed 15\mum~flux density. The total IR luminosity of this galaxy
is $L_{\rm IR}^{15}$. Using the same strategy, we computed $L_{\rm
IR}^{24}$. Lastly, we selected the SED which minimized the $\chi^2$ of
the observed 15 and 24\mum~flux densities with their associated error
bars and derived a third luminosity, $L_{\rm IR}^{15-24}$. The MIR
luminosities in the rest-frame of the galaxies (crosses proportional
to the filter width and error bar on the flux density) are compared to
the SEDs used to compute $L_{\rm IR}^{15-24}$ in
Fig.~\ref{FIG:all_seds} and their associated MIR luminosities (filled
circles).

The values of $L_{\rm IR}$ derived by the three techniques are equal
with a 1-$\sigma$ dispersion of only 20\,\%, which confirms the
robustness of this technique. Hence on average, the combination of
both instruments will affect the predicted $L_{\rm IR}$ values with a
20\,\% dispersion only with respect to the use of only one of the two
instruments. Hence the global shape of the template SEDs is such that
it can be used to predict with this accuracy the other MIR value,
suggesting that the SEDs did not vary very much since $z\sim$ 1.  In
the ``starburst'' regime, i.e. below $L_{\rm IR}=10^{11}$ L$_{\odot}$,
the template SEDs do not provide a good fit to the galaxy ``\#1''
($z=$ 0.4656, log[$L_{\rm IR}$/L$_{\odot}$]=10.7), which behaves very
similarly to NGC 7714 (Brandl et al. 2004) as shown in dashed line in
Fig.~\ref{FIG:all_seds}. The dispersion observed even among local galaxies,
for example in the MIR-FIR correlations already suggested that an
ideal library of template SEDs would have to include a variation of
shapes for each $L_{\rm IR}$ bin.

The bold SED in the middle of Fig.~\ref{FIG:all_seds} presents a
strong evidence for the presence of the 7.7\mum~PAH feature, band or
complex. The rest-frame 9\mum~luminosity of this 10$^{11.1}$
L$_{\odot}$ galaxy ($z\sim$ 0.7) is 1.8 times larger than the
14\mum~one, which is natural when PAHs are present but rules out a hot
dust continuum emission as the one locally observed in individual HII
regions (such as M17, Cesarsky et al. 1996b), dwarf galaxies (Thuan,
Sauvage \& Madden 1999) or even Compton-thick Seyfert 2's (such as NGC
1068, bold dashed line in the Fig.~\ref{FIG:all_seds}; Le Floc'h et
al. 2001).  The SEDs were sorted as a function of increasing redshift
in the Fig.~\ref{FIG:all_seds} in order to illustrate the effect of
the k-correction which acts as a low resolution spectrograph by
shifting the broadband ISOCAM and MIPS filters to lower wavelengths
with increasing redshift. The good quality of the fit was obtained
without allowing the luminosity of the template SEDs to vary which
illustrates the very good agreement with observations at
5-25\mum~rest-frame.

The templates from Chary \& Elbaz (2001) are therefore consistent with
the observed 24/15\mum~colors of distant sources considered as a
whole.  However it is worth mentioning that they do not provide a
unique solution for {\it every\,} galaxy of the sample, since some of
them can also be fitted by a feature-less continuum (e.g., ID~\#25
in Fig.\,2).  This issue is explored with more details in
Fig.~\ref{FIG:mean_sed} (similar to Fig.1 of Charmandaris et al. 2004) 
where we compare the colors derived from our
sources with what would be observed as a function of redshift for
different object types such as the Seyfert~2 prototype NGC~1068, two
enshrouded IR-luminous systems with deep silicate absorption (Arp\,220: Elbaz
et al. 2002; NGC\,4418: Spoon et al. 2001), M82 and three template SEDs
(as in Fig.\ref{FIG:all_seds}). The comparison shows
that galaxies with $\nu L_\nu(15)/\nu L_\nu(24) \leq 0.8$ can indeed be
explained by power-law spectra with no significant contribution
from PAHs.  This fitting
degeneracy   results in 
an additional uncertainty in the extrapolation to the total IR luminosity
(typically a factor of 2), and it
 also indicates that our approach is not sufficient in
itself to discreminate between starbursts and AGNs.
 Nonetheless, 
a non negligible fraction of sources at $z$\,$\sim$\,0.2-0.8 ($\sim$\,40-50\%)
exhibit  $\nu L_\nu(15)/\nu
L_\nu(24)$ colors larger than $\sim$\,1.
The latter  can not be  reproduced by SEDs without 
strong emission  features redshited 
in the ISOCAM band and boosting the flux at 15\mum.
As predicted by the starburst-dominated templates,
these colors are therefore 
the telltale signature for the presence of PAHs in the distant Universe.

%---------------------------------------------------------------------- 
%---------------------------------------------------------------------- 
%---------------------------------------------------------------------- 
\section{Conclusion} 
%\label{SECTION:conclusions}
%---------------------------------------------------------------------- 
%---------------------------------------------------------------------- 
%---------------------------------------------------------------------- 
We have shown that by combining ISO and {\it Spitzer} in the
photometric mode, it was possible to constrain the MIR SED of distant
galaxies too faint to be subject to direct spectroscopy in the
MIR. Deep images with both instruments and at similar depths (140 and
170\,$\mu$Jy at 15 and 24\mum) do detect the same objects which
illustrates the robustness of both instruments. The combination of 15
and 24\mum~flux densities measured with ISOCAM and MIPS was used to
constrain the rest-frame 5-25\mum~part of the SED of a sample of 16
LIRGs with $z\sim$ 0.1-1.2, taking advantage of the k-correction.
Similar studies with larger statistics will become feasible soon by combining 
MIPS 24\mum~with IRS 16\mum~peak-up imaging. Even though some sources 
were found to be either consistent with a starburst
or an AGN-dominated SED, a significant fraction of our sample shows clear
evidence for the presence of the broad bump associated to the
7.7\mum~PAH feature in emission and for the silicate feature in
absorption centered at 9.7\mum. Local template SEDs fitting the
correlations between MIR and FIR luminosities provide an overall good fit to
the distant LIRGs supporting the possibility to use them in models
fitting galaxy counts and deriving a cosmic star formation history
using a ``backward evolution'' of the MIR luminosity function.

\begin{acknowledgements}  
We are particularly grateful to B.Brandl and H.Spoon for providing us
with material used in this work. ELF thanks the MIPS project which is
 supported by NASA through the Jet Propulsion Laboratory, subcontract
 \#960785. DE thanks the support from the CNES.
\end{acknowledgements}  

%---------------------------------------------------------------------- 
%---------------------------------------------------------------------- 
%----------------------------------------------------------------------  

\end{document}